\documentstyle[prd,aps,epsf,floats,amsfonts,amssymb,amsmath]{revtex}

\begin{document}
 \renewcommand{\theequation}{\thesection.\arabic{equation}}
 \draft
  \title{Spin Flavor Conversion of Neutrinos in Loop Quantum Gravity
  }
\author{G. Lambiase$^{a,b}$}
\address{$^a$Dipartimento di Fisica "E.R. Caianiello"
 Universit\'a di Salerno, 84081 Baronissi (Sa), Italy.}
  \address{$^b$INFN - Gruppo Collegato di Salerno, Italy.}
\date{\today}
\maketitle
\begin{abstract}
Loop quantum gravity theory incorporates a new scale length ${\cal
L}$ which induces a Lorentz invariance breakdown. This scale can
be either an universal constant or can be fixed by the momentum of
particles (${\cal L}\sim p^{-1}$). Effects of the scale parameter
${\cal L}$ and helicity terms occurring in the dispersion relation
of fermions are reviewed in the framework of spin-flip conversion
of neutrino flavors.

\end{abstract}

\section{Introduction}
\setcounter{equation}{0}

Many attempts aimed to construct a theory of quantum gravity have
shown that spacetime might have a non trivial topology at the
Plank scale. The first to suggest that spacetime could have a
foam-like structure was Wheeler in his seminal paper
\cite{wheeler}. In the last years, the idea of quantum
fluctuations of spacetime background has received a growing
interest and today a large attention is addressed to study their
effects on physical phenomena (see, for example, Refs.
\cite{hawking80,hawking82,ellis84,ellis92,garay,ellis99,ford,ahluwalia}).

The difficulties to build up a complete theory of quantum gravity
has motivated the development of semiclassical approaches in which
a Lorentz invariance breakdown occurs at the effective theory
level. Deformations of the Lorentz invariance manifest by means of
a slight deviation from the standard dispersion relations of
particles propagating in the vacuum. Such modifications have been
suggested in the paper \cite{amelino}, as well as in the framework
of String Theory\footnote{In the framework of string theory, an
alternative approach has been proposed by Kostelecky and
collaborators. In their approach, even if the underlying theory
has Lorentz and CPT invariance symmetries, the vacuum solution of
the theory can spontaneously violate these symmetries
\cite{kostelecky}.} ($D$-branes) \cite{ellis} and Loop Quantum
Gravity \cite{gambini,pullinReview,alfaroPRL,alfaroPRD}. These
approaches foresee a dispersion relation in vacuo of particles of
the form (we shall use natural units $c=1=\hbar$)
 \begin{equation}\label{general}
 E^2\approx p^2+m^2+f\left(M,p\;l_{Pl}\right)\,,
 \end{equation}
where $f(x)$ is a model dependent function, $M$ fixes a
characteristic scale not necessarily determined by Planck length
$l_{Pl}\sim 10^{-19}$GeV$^{-1}$, and $p\,l_{Pl}\ll 1$. As a
consequence of Eq. (\ref{general}), the {\it quantum gravitational
medium} responds differently to the propagation of particles of
different energies.

Recently, Alfaro, Morales-T\'ecotl, and Urrutia (AMU) have written
a series of very important papers \cite{alfaroPRL,alfaroPRD} in
which it is developed a formalism based on loop quantum gravity
\cite{rovelli}. In this theory a new length scale ${\cal L}$
appears, ${\cal L}\gg l_{Pl}$, which separates the distances $d$
that manifest the quantum loop structure of space ($d\ll {\cal
L}$) from the continuous flat space ($d\gtrsim {\cal L}$). The new
characteristic scale can be either an universal constant, or it
can be fixed by the energy of the particle, ${\cal L}\sim E^{-1}$.
The latter case is called mobile scale.

In the AMU formalism, the dispersion relation of freely
propagating fermions is given by \cite{alfaroPRL}
\begin{equation}\label{disp-ferm}
  E_\pm^2=A_p^2p^2+\eta p^4 \pm 2\lambda p +m^2\,,
\end{equation}
where
\begin{equation}\label{coeff-ferm}
  A_p=1+\kappa_1\left(\frac{l_{Pl}}{{\cal L}}\right)\,, \qquad \eta=\kappa_3l_{Pl}^2\,, \qquad
  \lambda=\kappa_5 \frac{l_{Pl}}{2{\cal L}^2}\,.
\end{equation}
The $\pm$ signs stand for the helicity of the propagating
fermions, and $\kappa_i$ are unknown coefficients of the order
${\cal O}(1)$. For photons, the dispersion relation derived in AMU
theory is \cite{alfaroPRD}
\begin{equation}\label{disp-phot}
  \omega_\pm^2=k^2[A_\gamma^2\pm 2\theta_\gamma l_{Pl}k]\,,
\end{equation}
where
\begin{equation}\label{coeff-phot}
  A_\gamma=1+\kappa_\gamma\left(\frac{l_{Pl}}{{\cal
  L}}\right)^{2+2\Upsilon}\,.
\end{equation}
Again $\pm$ signs stand for the helicity dependence of photons in
the dispersion relation, $\Upsilon = -1/2, 0, 1/2, 1, \ldots$, and
$k_\gamma \sim {\cal O}(1)$. It is worth note that a similar
result has been obtained by Gambini and Pullin \cite{gambini} with
$A_\gamma=1$ and Ellis et al. \cite{ellis} with the difference
that the helicity dependence is absent.

On the experimental side, the present status eludes any
possibility to probe directly quantum gravity effects. It has been
suggested by Amelino-Camelia et al. \cite{amelino} that
$\gamma$-ray bursts might be a possible candidate to test the
theories of quantum gravity due to their peculiar physical
properties, i.e. the origin at cosmological distance and the high
energy, which might make them sensitive to the additional terms
occurring in (\ref{general}). Alfaro and Palma \cite{AlfaroPalma}
applies the AMU theory to the observed Greizen-Zatsepin-Kuz'min
(GZK) limit anomaly and to the so called TeV-$\gamma$ paradox,
i.e. the detection of high-energy photons with  a spectrum ranging
up to 24 TeV from Mk 501. Assuming that no anomalies there exist,
as recent works seem to indicate \cite{stecker}, they find that
the favorite range for ${\cal L}$ is
\begin{equation}\label{boundAP}
 4.6 \times 10^{-17}\mbox{eV}^{-1} \gtrsim {\cal L} \gtrsim 8.3\times
 10^{-18}\mbox{eV}^{-1}\,.
\end{equation}
Consequences of the dispersion relation (\ref{disp-ferm}) in other
contexts have been studied in Refs.
\cite{urrutiaCBR,major,gaetanoMPLA}.

The aim of this paper is to investigate some consequences of
fermion helicity terms appearing in the dispersion relation
(\ref{disp-ferm}). Such a kind of analysis has been performed for
photons (using Eq. (\ref{disp-phot})) by Gambini and Pullin
\cite{gambini}, and Gleiser and Kazameh \cite{gleiser}, which have
shown the emergence of a birefringence effect. Here we study the
effects of the helicity terms in spin flip conversion of neutrino
flavors. In Refs. \cite{alfaroPRL,gaetanoMPLA,hugo}, the
coefficients $k_i$ appearing in (\ref{disp-ferm}) are taken flavor
depending (hence responsible for the flavor mixing). To emphasize
the role of the helicity terms, we assume that $k_i$ are the same
for all species of neutrinos
($k_i^{(\nu_e)}=k_i^{(\nu_\mu)}=k_i^{(\nu_\tau)}=k_i$).

\section{Spin flavor oscillations in AMU Theory}
\setcounter{equation}{0}

Before to study the spin flavor oscillation of neutrinos, some
preliminary comments are in order. For the sake of convenience, we
shall write Eq. (\ref{disp-ferm}) in the relativistic regime
keeping out the relevant terms for our considerations. In this
approximation the helicity operator is equal, up to the factor
$(m/E)^2\ll 1$, to the chiral operator \cite{zuber}. We then get
\begin{equation}\label{disp-helicity}
  E_{L, R}\simeq p+\frac{m^2}{2p}\mp \lambda +F(p, {\cal L},
  l_{Pl}, k_i)\,,
\end{equation}
where the function $F(x)$ contains all remaining contributions in
(\ref{disp-ferm}), and it is diagonal in the chiral basis. The
terms $\pm \lambda$ change the effective energy of neutrinos
depending if they are left-handed or right handed. This is a very
peculiar aspect of the AMU theory since it gives rise, as we will
see, to the occurrence of a resonance condition in the analysis of
spin flavor oscillations.

As shown by Mikheyev, Smirnov, and Wolfeinstein \cite{MSW}, when
neutrinos propagate in a medium, their energy is shifted owing to
the weak interaction with the background matter. The elastic
scattering through charged current interaction gives the energy
contribution $2\sqrt{2}\, G_Fn_e$, whereas the neutral current
interaction gives $\sqrt{2}\, G_Fn_n$, where $G_F$ is the Fermi
constant, and $n_e$ ($n_n$) is the electron (neutron) density. The
net contribution to the energy is therefore
$\sqrt{2}G_F(2n_e-n_n)\lesssim \sqrt{2}G_Fn_e$ ($n_n\lesssim
n_e$). Notice that for the Sun, a reasonable profile for $n_e$ is
$n_e(r)=n_0e^{-10.5 r/R_\odot}$, where $n_0=85 N_A$cm$^{-3}$,
$N_A$ is the Avogadro number \cite{bahcall,raffelt}. At $r=0$, one
gets $\sqrt{2}G_Fn_e(0)\sim 10^{-12}$eV \cite{bilenky}. For the
Earth, $\sqrt{2}G_Fn_e(0)\sim 10^{-14}$eV \cite{bilenky}. Besides,
in order to have non vanishing off-diagonal elements in the
effective Hamiltonian, we shall take into account of the
interaction between neutrinos and an external magnetic field
\cite{okun}
\begin{equation}\label{magn-int}
  {\cal L}_{int}={\bar \psi}\mu\sigma^{\mu\nu}F_{\mu\nu}\psi\,,
\end{equation}
where $\mu$ is the magnetic momentum of the neutrino, $F_{\mu\nu}$
is the electro-magnetic field tensor, and
$\sigma^{\mu\nu}=\frac{1}{4}[\gamma^{\mu}, \gamma^{\nu}]$.

Collecting all that, the equation of evolution describing the
conversion between two neutrino flavors is therefore
\cite{raffelt,PIN}
 \begin{equation}\label{11}
i\frac{d}{d r}\left(\begin{array}{c}
                           \nu_{eL} \\
                             \nu_{\mu L} \\
                              \nu_{eR} \\
                           \nu_{\mu R}\end{array}\right)={\cal H}\left(\begin{array}{c}
                                                            \nu_{eL} \\
                             \nu_{\mu L} \\
                              \nu_{eR} \\
                           \nu_{\mu R}\end{array}\right)\,,
 \end{equation}
where, in the chiral base, the matrix ${\cal H}$ is the effective
Hamiltonian defined as \cite{MSW,bilenky,raffelt,palm}
\begin{equation}\label{12}
{\cal H}=\left[\begin{array}{cccc}
 \displaystyle{-\frac{\Delta m^2}{4p}}\cos 2\theta- \lambda +\sqrt{2}G_Fn_e
         &  \displaystyle{\frac{\Delta m^2}{4p}}\sin 2\theta & \mu_{ee} B & \mu B \vspace{0.05in} \\
 \displaystyle{\frac{\Delta m^2}{4p}}\,\sin 2\theta  &
 \displaystyle{\frac{\Delta m^2}{4p}}\cos 2\theta - \lambda  +\sqrt{2}G_Fn_e & \mu B & \mu_{\mu\mu} B \vspace{0.05in}\\
 \mu_{ee} B & \mu B & -\displaystyle{\frac{\Delta m^2}{4p}}\cos 2\theta + \lambda &
                        \displaystyle{\frac{\Delta m^2}{4p}}\sin 2\theta \vspace{0.05in}\\
 \mu B & \mu_{\mu\mu} B & \displaystyle{\frac{\Delta m^2}{4p}}\sin 2\theta &
                        \displaystyle{\frac{\Delta m^2}{4p}}\cos 2\theta +\lambda
\end{array}\right]\,,
\end{equation}
up to terms proportional to identity matrix. Here $\Delta
m^2\equiv m_2^2-m_1^2$, $B$ is the magnetic field (the electric
field is zero), and $\mu=\mu_{e\mu}$. In the more general case the
effective Hamiltonian ${\cal H}$ does contain also a gravitational
field term which is proportional to
 \[
 A_G^\mu=\frac{1}{4}e^\mu_a\varepsilon^{abcd}(e_{b\nu; \sigma}-
 e_{b \sigma; \nu})e_c^\nu e_d^\sigma
 \]
where $e_a^\mu$ are the vierbein fields and ";" means the
covariant derivative. Such a term vanishes for Schwarzschild-like
geometry due to the spherical symmetry, so that no gravitational
contributions arise in our analysis. We restrict to flavors $\nu_e
- \nu_\mu$, but obviously this analysis works also for different
neutrino flavors ($\nu_\mu -\nu_\tau$, $\nu_e-\nu_\tau$).

The terms giving the resonant condition of flavor spin-flip are
the diagonal elements of the effective Hamiltonian (\ref{12}).
Hence
\begin{eqnarray}
 \nu_{eL}\to \nu_{\mu R} & \qquad \qquad & 2\lambda +\frac{\Delta m^2}{2p}\, \cos 2 \theta
 -\sqrt{2}G_Fn_e(r_{res})=0\,,
 \label{res1} \\
   \nu_{\mu L}\to \nu_{eR} & \quad \qquad & 2\lambda -\frac{\Delta m^2}{2p}\, \cos 2 \theta
   -\sqrt{2}G_Fn_e(r_{res})=0\,.
 \label{res2}
\end{eqnarray}
In both cases, we only have transitions from left-handed to
right-handed neutrinos, the latter being sterile neutrinos {\it do
not have} electroweak interaction with matter. Clearly, the
resonant transitions do not occur simultaneously. From now on we
assume that the neutrino mass hierarchy is $m_1>m_2$ ($\Delta
m^2=-|\Delta m^2|$). Notice that expressions similar to Eqs.
(\ref{res1}) and (\ref{res2}) hold also for Majorana neutrinos,
with the difference that $\nu_{eL}, \nu_{\mu L}\to \nu_{e},
\nu_{\mu}$ and $\nu_{eR}, \nu_{\mu R}\to {\bar \nu}_{e}, {\bar
\nu}_{\mu}$ in Eq. (\ref{11}) \cite{raffelt,wudka2}.

Let us analyze the $\nu_{e L}\to \nu_{\mu R}$ transition. From Eq.
(\ref{res1}), one gets ($E\sim p$)
\begin{equation}\label{bound}
  \vert \cos 2\theta \vert \sim \frac{2E[2\lambda -
  \sqrt{2}G_Fn_e(r_{res})]}{|\Delta m^2|}\lesssim 1\,.
\end{equation}
The transition between the two flavors $\nu_{e L}\to \nu_{\mu R}$
is determined by defining the effective mixing angle ${\tilde
\theta}$ which diagonalizes the corresponding submatrix in
(\ref{12})
 \begin{equation}\label{14}
 \tan 2\tilde{\theta}(r)=\frac{4\mu B(r) E}{|\Delta m^2|\cos
 2\theta-4E\lambda+2\sqrt{2}G_FEn_e(r)}\,.
 \end{equation}
The resonance condition, for which neutrino oscillations are
appreciably enhanced, occurs when the denominator of (\ref{14})
vanishes. When $\lambda=0$ there is no resonance unless the
neutrino mass hierarchy is inverted ($m_2>m_1$ and the usual
resonance condition is restored). Flavor eigenstates and mass
eigenstates are now related by the effective mixing angle
 \[
 |\nu_e\rangle=\cos {\tilde \theta}|\nu_1\rangle +\sin {\tilde
 \theta}|\nu_2\rangle\,, \qquad
 |\nu_\mu\rangle=-\sin {\tilde \theta}|\nu_1\rangle +\cos {\tilde
 \theta}|\nu_2\rangle
 \]
For adiabatic variation of the magnetic
field\footnote{Unfortunately, the profile of magnetic fields in
the core, radiation zone or convection zone of Sun is little
known, except that they may be quite large. Similarly for Earth.
In the case in which the magnetic field is homogeneous (constant),
the conversion probability is (see, for example, the paper by Lim
and Marciano in Ref. \cite{okun})
\begin{equation}\label{P1a}
   P_{\nu_{eL}\to \nu_{\mu R}}=\frac{(2\mu B)^2}{(2\mu B)^2+
   (\frac{|\Delta m^2|}{2E}\cos 2\theta-
   2\lambda+\sqrt{2}G_Fn_e)^2}\sin^2 \alpha\,,
\end{equation}
where
 \[
 \alpha = \sqrt{\left(\frac{|\Delta m^2|}{2E}\cos 2\theta
 -2\lambda+\sqrt{2}G_Fn_e\right)^2
 +\left(2\mu B\right)^2}\,\,\Delta r
 \]
and $\Delta r$ is the distance travelled by neutrinos. At the
resonance it becomes
\begin{equation}\label{P2}
 P_{\nu_{eL}\to \nu_{\mu R}}=\sin^2 \mu B \Delta r\,.
\end{equation}
The neutrino magnetic momentum is $\mu\sim 10^{-11}\mu_B\sim
6\times 10^{-16}$eVT$^{-1}$ \cite{raffelt}. Taking the magnetic
field of Earth $B_\oplus\approx 5\times 10^{-5}$T, constant over
the region $\Delta r\sim 2 R_\oplus\sim 1.2 \times 10^7$m, it
follows that the transition probability is $P_{\nu_{eL}\to
\nu_{\mu R}}\sim 3\times 10^{-4}$, a value very small for giving
an appreciable spin flavor rotation effect. For the Sun, the
superficial magnetic field $B_\odot\sim 10^{-1}$T \cite{raffelt},
and $\Delta r\sim R_\odot\sim 7\times 10^8$m give $P_{\nu_{eL}\to
\nu_{\mu R}}\sim 0.04$. On the other hand, in the convective zone
$B_\odot\sim 10$T \cite{raffelt}, thus $P_{\nu_{eL}\to \nu_{\mu
R}}\sim 0.65$.} and electron density, the conversion probability
is \cite{parke,haxton}
 \begin{equation}\label{19}
 P_{\nu_{eL}\to \nu_{\mu R}}=\frac{1}{2}(1-\cos 2\tilde{\theta}\cos 2\theta)\,{,}
 \end{equation}
and at the resonance (${\tilde \theta}=\pi/4$) it assumes the
maximum value
 \begin{equation}\label{prob-res}
 P_{\nu_{eL}\to \nu_{\mu R}}=\frac{1}{2}\,.
 \end{equation}
Let us now use Eq. (\ref{bound}) to determine the bounds on the
scale parameter ${\cal L}$ and the coefficient $k_5$. We shall
consider the cases in which $i)$ ${\cal L}$ is an universal
constant, $ii)$ ${\cal L}\sim 10^{-18}$eV$^{-1}$ as recent results
seem to suggest \cite{AlfaroPalma,gaetanoMPLA,hugo,Palma2}, and
$iii)$ ${\cal L}\sim p^{-1}$ (mobile scale).

\subsection{Solar Neutrinos}

Due to the fact that the matter densities $n_e$ varies with the
distance $r$, as well as the magnetic field $B(r)$, the mass
eigenstates evolve adiabatically only if the adiabaticity
condition is satisfied at the resonance
\cite{bilenky,raffelt,palm}, that is
 \begin{equation}\label{gamma}
 \gamma \equiv \left\vert\frac{a_1-a_2}{d{\tilde \theta}/dr}\right\vert_{res}\gg
 1\,,
 \end{equation}
where the effective mixing angle is define by (\ref{14}) and
$a_i$, $i=1, 2$, are the eigenvalues of the submatrix of evolution
Eq. (\ref{12}) corresponding to $\nu_{eL}-\nu_{\mu R}$ transition
\begin{equation}\label{lambda12}
  a_{1,2}=\frac{1}{2}\left\{
  \sqrt{2}\, G_Fn_e \pm
  \sqrt{\left(\frac{|\Delta m^2|}{2E}-2\lambda+\sqrt{2}\,
  G_Fn_e\right)^2+4\mu^2|B_{\odot}|^2}
  \right\}\,.
\end{equation}
The transition probability (\ref{19}) is generalized by the
formula \cite{parke}
 \begin{equation}\label{parke}
 P_{\nu_{eL}\to \nu_{\mu R}}=\frac{1}{2}-\left(\frac{1}{2}-
 P_c\right)\cos 2\tilde{\theta}\cos 2\theta\,{,}
 \end{equation}
where $P_c=\exp[{-2\pi\gamma \sin^2{\tilde \theta}\cos{\tilde
\theta}/\sin^22{\tilde \theta}}]$ \cite{bilenky}. If $\gamma \gg
1$ realizes, then the formula (\ref{19}) is recovered.

Eq. (\ref{14}) allows to calculate $(d{\tilde \theta}/dr)_{res}$,
and by using (\ref{lambda12}), the adiabaticity condition
(\ref{gamma}) assumes the form
\begin{equation}\label{adiabatic1}
 \gamma^{-1}= \left(\frac{\sqrt{2}G_F}{8\mu^2 |B_\odot|^2}\left|
  \frac{dn_e}{dr}\right|  \right)_{res} \ll 1\,.
\end{equation}
The term containing $dB/dr$ is absent since at the resonance its
coefficient vanishes. Being $B_\odot\gtrsim 10$T in the solar
convective zone \cite{raffelt}, Eq. (\ref{adiabatic1}) reduces to
\begin{equation}\label{adiabatic2}
 e^{-10.5 r_{res}/R_{\odot}} \ll 0.1 \,,
\end{equation}
which is fulfilled by $r_{res}\gtrsim 0.4 R_{\odot}$.
Eq. (\ref{adiabatic2}) then assures that the mass eigenstates of
neutrinos produced in the Sun evolve always adiabatically.

{} From (\ref{bound}) it follows the upper bound on $\lambda$
 \begin{equation}\label{upperSun}
 \lambda \lesssim \frac{1}{2}\left[10^{-12}e^{-10.5 r_{res}/R_\odot}\mbox{eV}+
 \frac{|\Delta m^2|}{2E}\right]\,.
 \end{equation}
The best fit for neutrino oscillations induced by MSW effect is
given by the following (experimental) values of the $\Delta m^2
-\sin^2 2\theta$ parameters \cite{bilenky}
 \[
 \begin{array}{ccc}
   |\Delta m^2|\sim (3\div 10)\times 10^{-6}\mbox{eV}^2 &
   \quad \sin^22\theta\sim (0.6\div 1.3)\times 10^{-2} &\quad \mbox{SMA} \\
   |\Delta m^2|\sim (1\div 20)\times 10^{-5}\mbox{eV}^2 & \quad \sin^22\theta\sim 0.5\div 0.9
   &\quad   \mbox{LMA} \
 \end{array}
 \]
where SMA and LMA stand for Small ($\sin^2 2\theta \ll 1$) and
Large ($\sin^2 2\theta \lesssim 1$) Mixing Angle solution,
respectively.

In what follows, estimations are carried out for solar neutrinos
with the maximum energy $E\sim 15$MeV, produced by $^{8}B$ and
$hep$ reactions \cite{bilenky}. It is worth to point out that even
though the flux of $^8B$ neutrinos produced in the Sun is much
smaller than the fluxes of $pp$, $^7Be$, and $pep$ neutrinos,
$^{8}B$ neutrinos give the major contribution to the event rates
of experiments with a high energy detection threshold
\cite{bilenky}. In fact, the results of Kamiokande and
SuperKamiokande experiments are usually presented in terms of the
measured flux of $^8B$ neutrinos \cite{bilenky}.

The resonance occurring at $r_{res}\gtrsim 0.4 R_{\odot}$ implies
that the background matter density reduces and the dominant term
in (\ref{upperSun}) is the massive one. For $k_5\sim {\cal O}(1)$,
the lower bound on ${\cal L}$ is ${\cal L}\gtrsim
10^{-6}$eV$^{-1}$, which is nearly in the range of nuclear physics
so that it can be discarded since no evidence of loop structure
occurred at this scale. If ${\cal L}\sim 10^{-18}$eV$^{-1}$, then
the upper bound on $k_5$ is $k_5 < 10^{-25}$, which is a weak
bound since the coefficient $k_5$ is expected to be of the order
${\cal O}(1)$. Finally, in the case of mobile scale, one gets $k_5
\sim {\cal O}(1) \div {\cal O}(10)$ for the SMA solution, and $k_5
< {\cal O}(10) \div {\cal O}(10^2)$ for the LMA solution, that are
the expected order of magnitude for $k_5$. Such ranges increase
for neutrinos with energies $\lesssim 10$MeV (as, for instance,
for $pp$, $^7Be$, $pep$ neutrinos).


Let us finally discuss the case of degenerate or massless
neutrinos (or $\theta=\pi/4$). The effective mixing angle then
reduces to
 \[
 |\tan 2{\tilde \theta}| \sim \frac{\mu B_\odot}{\lambda}\,,
 \]
where we have neglected the weak interaction of neutrinos with the
background matter ($n_e\approx 0$). We consider neutrinos
propagating in regions far from the Sun core, where the
superficial magnetic field of the Sun is $B_\odot\sim 10^{-1}$T
\cite{barbieri}. Being the neutrino magnetic momentum $\mu\sim
10^{-11}\mu_B\sim 6\times 10^{-16}$eVT$^{-1}$ \cite{raffelt}, it
follows that the magnetic energy of neutrinos is $\mu B_\odot\sim
6\times 10^{-17}$eV. Since $k_5\sim {\cal O}(1)$ and taking ${\cal
L}\sim p^{-1}$, with $p\sim $MeV, one gets $|\tan 2{\tilde
\theta}|\sim 1$, i.e. $\sin^2 2 {\tilde \theta}\sim 0.5$. The
transition probability (\ref{P1a}) gives $P_{\nu_{eL}\to \nu_{\mu
R}}=0.16$.

\subsection{Atmospheric Neutrinos}

Strong evidence in favor of atmospheric neutrino oscillations come
from SK experiment \cite{SK}, and the relevant values of $|\Delta
m^2| - \sin^2 2\theta$ parameters are ($\nu_\mu$-$\nu_\tau$)
\cite{bilenky,fogli}
\begin{equation}\label{bestfit}
  |\Delta m^2|\sim 5\times 10^{-3}\mbox{eV}^2\,, \qquad \sin^2 2\theta
  \sim 1\,,
\end{equation}
which implies $\cos 2\theta < 1$. From Eq. (\ref{bound}) it
follows ($E\sim $GeV) ${\cal L}> 10^{-9}$eV$^{-1}$ as $k_5\sim
{\cal O}(1)$, which has to be discarded, as discussed before. If
the scale length is fixed by ${\cal L}\sim 10^{-18}$eV$^{-1}$,
then the parameter $k_5$ is bounded by $k_5\ll 10^{-16}$. Finally,
in the case of mobile scale, one gets
 \begin{equation}\label{kappaAtm}
 k_5 < \frac{|\Delta m^2|}{4E^3 l_{Pl}}\sim 10^{-2}\,.
 \end{equation}
For neutrinos with sub-GeV energies, $E\gtrsim 100$MeV, the
coefficient $k_5$ in (\ref{kappaAtm}) is bounded from above by
$k_5\lesssim 10$, which agrees with expected order of $k_5\sim
{\cal O}(1)$.

The channel of oscillation $\nu_\mu \to \nu_e$ in the accelerator
LSND experiment,
  fits experimental data for the following region of parameters \cite{LSND}
\begin{equation}\label{par-LSND}
 |\Delta m^2|\sim \mbox{eV}^2\,, \qquad \sin^2 2\theta
  \approx (0.2 \div 3)\times 10^{-3}\,,
\end{equation}
thus $\cos 2\theta \approx 1$. The maximum neutrino energy is
taken of the order $E\sim 10$MeV. As $k_5\sim 1$, we find ${\cal
L}\sim 10^{-9}$eV$^{-1}$. For ${\cal L}\sim 10^{-18}$eV$^{-1}$, it
follows $k_5\sim 10^{-17}$. Finally, for ${\cal L}\sim E^{-1}$,
the parameter $k_5$ is given by $k_5\sim 10^4$. We have to point
out that LSND experiments still remains unconfirmed, contrarily to
the solar and atmospheric experiments which have been confirmed by
other experiments.

\vspace{0.5cm}

The above results show that the helicity terms appearing in the
dispersion relation (\ref{disp-ferm}) are the responsible for the
occurrence of the resonance condition in the spin-flip conversion
of neutrino flavors, whereas the background matter effects are
negligible (or are absent). The estimations provide a narrow
region of solar and atmospheric neutrino energies in which
experiments might seek effects induced by helicity terms occurring
in loop quantum gravity, as suggested by AMU's theory.

\section{Conclusion}
Despite the concrete difficulties to probe quantum gravity effects
occurring at the Planck scale, it is an intriguing suggestion that
quantum gravity should predict a slight departure from Lorentz's
invariance, which manifests in a deformation of the dispersion
relations of photons and fermions. Such results have been indeed
suggested in loop quantum gravity
\cite{gambini,alfaroPRL,alfaroPRD} (and string theory
\cite{kostelecky,ellis}). This approach is endowed with a scale
length characterizing the scale on which new effects are non
trivial. The estimation of the order of magnitude of the new scale
length and of parameters entering in the dispersion relations is
certainly of current interest, as well as different scenarios
where these effects become testable \cite{AlfaroPalma,neweffects}.

In this paper we have determined bounds on ${\cal L}$ and $k_5$
(occurring in Eq. (\ref{disp-ferm})) in the context of neutrinos
oscillations. In the case of solar and atmospheric neutrino
oscillations, the analysis performed suggests that the mobile
scale ${\cal L}\sim p^{-1}$ is the favorite one, with $k_5
\lesssim {\cal O}(1)\div {\cal O}(10)$, as expected. Thus,
neutrino oscillations, seem to be very promising candidates (see
also Refs. \cite{alfaroPRL,gaetanoMPLA,gleiser,hugo,foffa}) for
bringing quantum gravity effects predicted by AMU theory to the
realm of experimental evidence.



\end{document}